\documentclass[sigconf]{acmart}

\setcopyright{none} 
\renewcommand\footnotetextcopyrightpermission[1]{} 
\usepackage{booktabs} 
\usepackage{graphicx}
\usepackage{multirow}
\usepackage{array}
\usepackage{amsmath}
\usepackage{paralist}
\usepackage{subcaption}
\usepackage[stable]{footmisc}

\newcounter{inlinecounter}
\newcommand{\icstart}{\setcounter{inlinecounter}{0}}
\newcommand{\ic}{({\protect\refstepcounter{inlinecounter}\theinlinecounter)~}}

\newcommand{\Generator}{{\sf $G$}}
\newcommand{\Discriminator}{{\sf $D$}}

\newcommand{\usenix}{FLA}

\begin{document}
\title{PassGAN: A Deep Learning Approach for Password Guessing} 
\titlenote{This is an extended version of the paper~\cite{Hitaj2018PassGAN}, which appeared in NeurIPS 2018 Workshop on Security in Machine Learning (SecML'18).}

\author{Briland Hitaj}
\affiliation{
  \institution{Stevens Institute of Technology}
}
\email{bhitaj@stevens.edu}

\author{Paolo Gasti}
\affiliation{
  \institution{New York Institute of Technology}
}
\email{pgasti@nyit.edu}

\author{Giuseppe Ateniese}
\affiliation{
  \institution{Stevens Institute of Technology}
}
\email{gatenies@stevens.edu}

\author{Fernando Perez-Cruz}
\affiliation{
  \institution{Swiss Data Science Center, (ETH Zurich and EPFL)}
}
\email{fernando.perezcruz@sdsc.ethz.ch}

\renewcommand{\shortauthors}{Hitaj et al.}

\begin{abstract}

State-of-the-art password guessing tools, such as HashCat and John the Ripper,
enable users to check billions of passwords per second against password hashes.
In addition to performing straightforward dictionary attacks, these tools can
expand password dictionaries using password generation rules, such as
concatenation of words (e.g., ``password123456'') and {\em leet speak} (e.g.,
``password'' becomes ``p4s5w0rd''). Although these rules work well in practice,
expanding them to model further passwords is a laborious task that requires
specialized expertise.

To address this issue, in this paper we introduce PassGAN, a novel approach that
replaces human-generated password rules with theory-grounded machine learning
algorithms. Instead of relying on manual password analysis, PassGAN uses a
Generative Adversarial Network (GAN) to autonomously learn the distribution of
real passwords from actual password leaks, and to generate high-quality password 
guesses.
Our experiments show that this approach is very promising. When we evaluated
PassGAN on two large password datasets, we were able to surpass rule-based and
state-of-the-art machine learning password guessing tools. However, in contrast
with the other tools, PassGAN achieved this result without any a-priori
knowledge on passwords or common password structures.
Additionally, when we combined the output of PassGAN with the output of HashCat,
we were able to match 51\%-73\% more passwords than with HashCat alone. This is
remarkable, because it shows that PassGAN can autonomously extract a
considerable number of password properties that current state-of-the art rules
do not encode.

\end{abstract}

%
%
%


\maketitle


\section{Introduction}

Passwords are the most popular authentication method, mainly because they are
easy to implement, require no special hardware or software, and are familiar to
users and developers~\cite{huntessay}.
Unfortunately, multiple password database leaks have shown that users tend to
choose easy-to-guess
passwords~\cite{dell2010password,durmuth2015omen,ma2014study}, primarily
composed of common strings (e.g., {\tt password}, {\tt 123456}, {\tt iloveyou}),
and variants thereof.

Password guessing tools provide a valuable tool for identifying weak passwords when they are stored in hashed
form~\cite{percival2016scrypt,provos1999bcrypt}. 
The effectiveness of password guessing software relies on the ability to quickly
test a large number of highly likely passwords against each password hash.
Instead of exhaustively trying all possible character combinations, password
guessing tools use words from dictionaries and previous password leaks as
candidate passwords.
State-of-the-art password guessing tools, such as John the Ripper~\cite{jtr} and
HashCat~\cite{hashcat}, take this approach one step further by defining
heuristics for password transformations, which include combinations of multiple
words (e.g., {\tt iloveyou123456}), mixed letter case (e.g., {\tt iLoVeyOu}),
and {\em leet speak} (e.g., {\tt il0v3you}). These heuristics, in conjunction
with Markov models,
allow John the Ripper and HashCat to generate a large number of {\em new}~highly
likely passwords.

While these heuristics are reasonably successful in practice, they are ad-hoc
and based on intuitions on how users choose passwords, rather than being constructed from a principled analysis of large password datasets. For this
reason, each technique is ultimately limited to capturing a specific subset of
the password space which depends upon the intuition behind that technique. Further,
developing and testing new rules and heuristics is a time-consuming task that
requires specialized expertise, and therefore has limited scalability.

\subsection{Our Approach}
To address these shortcomings, in this paper we propose to replace rule-based
password guessing, as well as password guessing based on simple data-driven
techniques such as Markov models, with a novel approach based on deep learning.
At its core, our idea is to train a neural network to determine autonomously
password characteristics and structures, and to leverage this knowledge to
generate new samples that follow the same distribution. We hypothesize that deep
neural networks are expressive enough to capture a large variety of properties
and structures that describe the majority of user-chosen passwords; at the same
time, neural networks can be trained without any a priori knowledge or
an assumption of such properties and structures. This is in stark contrast with
current approaches such as Markov models (which implicitly assume that all
relevant password characteristics can be defined in terms of $n$-grams), and
rule-based approaches (which can guess only passwords that match with the
available rules). As a result, samples generated using a neural network are not
limited to a particular subset of the password space. Instead, neural networks
can autonomously encode a wide range of password-guessing knowledge that
includes and surpasses what is captured in human-generated rules and Markovian
password generation processes.

To test this hypothesis, in this paper we introduce PassGAN, a new approach for
generating password guesses based on deep learning and Generative Adversarial
Networks (GANs)~\cite{goodfellow2014generative}. GANs are 
recently-introduced
machine-learning tools designed to perform density estimation in
high-dimensional spaces~\cite{goodfellow2014generative}. 
GANs perform implicit generative modeling by training a deep neural network
architecture that is fed a simple random distribution (e.g., Gaussian or
uniform) and by generating samples that follow the distribution of the available
data. In a way, they implicitly model the inverse of the cumulative distribution with a deep neural network, i.e., $\mathbf{x}=F_{\theta}^{-1}(s)$ where
$s$ is a uniformly distributed random variable. To learn the generative model, GANs
use a cat-and-mouse game, in which a deep generative network (\Generator{}) tries to
mimic the underlying distribution of the samples, while a discriminative deep
neural network (\Discriminator{}) tries to distinguish between the original training samples
(i.e., ``true samples'') and the samples generated by \Generator{} (i.e., ``fake
samples''). This adversarial procedure forces \Discriminator{} to leak the relevant
information about the training data. This information helps \Generator{} to adequately reproduce the original data distribution.
PassGAN leverages this technique to generate new password guesses. We train \Discriminator{}
using a list of leaked passwords (real samples).
At each iteration,
the output of PassGAN (fake samples) gets closer to the distribution of
passwords in the original leak, and therefore more likely to match real users'
passwords. To the best of our knowledge, this work is the first to use GANs for
this purpose.

\subsection{Contributions}
PassGAN represents a principled and theory-grounded take on the generation of
password guesses. We explore and evaluate different neural network
configurations, parameters, and training procedures, to identify the appropriate
balance between {\em learning} and {\em overfitting}, and report our results.
Specifically, our contributions are as follows:

\begin{enumerate}

\item 
We show that a GAN can generate high-quality password
guesses. Our GAN is trained on a portion of the RockYou dataset~\cite{rockyou_dataset}, and tested on two different datasets: (1) another (distinct) subset of the RockYou dataset; and (2) a dataset of leaked passwords from LinkedIn~\cite{linkedin_dataset}. 
In our experiments, we were able to match 1,350,178 (43.6\%) \emph{unique} passwords out of 3,094,199 passwords from the RockYou dataset, and 10,478,322 (24.2\%) unique passwords out of 43,354,871 passwords from the LinkedIn dataset. To quantify the ability of PassGAN to generate new passwords, we removed from the testing set all passwords that were present also in the training set. This resulted in testing sets of size 1,978,367 and 40,593,536 for RockYou and LinkedIn, respectively. In this setting, PassGAN was able to match 676,439 (34.6\%) samples in the RockYou testing set and 8,878,284 (34.2\%) samples in the LinkedIn set. 
Moreover, the overwhelming majority of
passwords generated by PassGAN that did not match the testing sets still ``looked
like'' human-generated passwords, and thus could potentially match real user
accounts not considered in our experiments.

\item We show that PassGAN is competitive with state-of-the-art password
generation rules. Even though these rules were specially tuned for the
datasets used in our evaluation, the quality of PassGAN's output was comparable
to 
that of password rules.
\item With password generation rules, the number of unique passwords that can be
generated is defined by the number of rules and by the size of the password
dataset used to instantiate them. In contrast, PassGAN can output a practically
unbounded number of password guesses. Crucially, our experiments show that with
PassGAN the number of matches increases steadily with the number of passwords
generated. This is important because it shows that the output of PassGAN is not
restricted to a small subset of the password space.
As a result, in our experiments, PassGAN was able to eventually guess more
passwords than any of the other tools, even though all tools were trained on the
same password dataset. However, this result required PassGAN to generate a more significant
number of passwords with PassGAN than with the other~tools. 
\item PassGAN is competitive with the current state of the art password
guessing algorithms based on deep neural networks~\cite{MelicherPassGuessNN}. Our
results show that PassGAN essentially matches the performance of Melicher et
al.~\cite{MelicherPassGuessNN} (indicated as \usenix~in the rest of the paper).
\item We show that PassGAN can be effectively used to \emph{augment} 
password generation rules. In our experiments, PassGAN matched passwords that were not
generated by any password rule. When we combined the output of PassGAN with the
output of HashCat, we were able to guess between 51\% and 73\% additional unique
passwords compared to HashCat alone.

\end{enumerate}

We consider this work as the first step toward a fully automated generation of
high-quality password guesses.
Currently, there is a tradeoff between the benefits of PassGAN (i.e.,
expressiveness, generality, and ability to autonomously learn from samples), and
its cost in terms of output size, compared to rule-based approaches. While
rule-based password guessing tools can generate a significant number of
matches within a remarkably small number of attempts,
PassGAN must output a more significant number of passwords to achieve the same
result. We argue that this is, in practice, not an issue because: \icstart \ic
password guessing tools can be easily combined in such a way that once a fast
tool (e.g., HashCat) has exhausted its attempts, a more comprehensive one (such
as PassGAN) can continue to generate new matches; and \ic the cost of storage
has been steadily decreasing for decades, to the point that a cheap 
8~TB hard drive can store roughly $10^{12}$ password guesses. 
As such, password
generation can be treated as an offline process.

For these reasons, we believe that meaningful comparisons
between password guessing techniques should primarily focus on the
number of matches that each technique can generate, rather than
{\em how quickly} these matches are generated.

We argue that this work is relevant, important, and timely. {\em Relevant},
because despite numerous
alternatives~\cite{googleabacus,sitova2016hmog,frank2013touchalytics,duc1999face,zhong2012keystroke},
we see little evidence that passwords will be replaced any time soon. {\em
Important}, because establishing the limits of password guessing---and better
understanding how guessable real-world passwords are---will help make
password-based systems more secure. And {\em timely}, because recent leaks
containing hundreds of millions of passwords~\cite{yahoo_leak} provide a
formidable source of data for attackers to compromise systems, and for system
administrators to re-evaluate password policies.

\subsection{PassGAN in the Media and in Academia}

The ability of PassGAN to autonomously learn characteristics and patterns
constituting a password, much like DeepMinds' AlphaGo's ability to autonomously
learn the game of Go~\cite{alphago}, drew significant attention from several
media outlets. For instance, PassGAN has been reported in articles from Science
Magazine~\cite{sciencemagnews}, The Register~\cite{theregisternews}, Inverse
\cite{inversenews}, Threatpost~\cite{threatpostnews}, Dark Reading
\cite{darkreadingnews}, Technology Review News~\cite{technologyreviewnews},
Sensors Online~\cite{sensorsonlinenews}, and others~\cite{wsj2018, stevensnews, focusitnews, techtargetnews, hackernoonnews, stuckintrafficYouTube}. Further,
PassGAN was selected by Dark Reading as one of \emph{the coolest hacks of
2017}~\cite{darkreadingranking}. 

UC Berkeley has included PassGAN as reading
material in their graduate-level course titled {\em Special Topics in Deep
Learning}~\cite{berkeleyclasses}. 

\subsection{Changes with Respect to an Earlier Version of this Paper}
\label{ss:changes}

This paper updates and extends an earlier version of our
work~\cite{DBLP:journals/corr/abs-1709-00440}. The differences between the two
versions of the paper can be summarized as follows: \icstart \ic we identified
an issue with the PassGAN implementation used
in~\cite{DBLP:journals/corr/abs-1709-00440}, which led to a substantial
decrease in the number of unique passwords generated. We corrected
this issue and, as a result, in this paper we report a rate of generation of
unique passwords roughly four times higher than in our earlier work; and
\ic in the updated paper, we compare PassGAN with state-of-the-art password
guessing based on Markov Models, and with the work on neural-network (RNN) by
Melicher et al.~\cite{MelicherPassGuessNN}, in addition to John the Ripper and HashCat.

\subsection{Organization} The rest of this paper is organized as follows. In
Section~\ref{sec:background}, we briefly overview GANs,
password guessing, and provide a summary of the relevant state of the art.
Section~\ref{sec:sys_design} discusses the architectural and training choices
for the GAN used to instantiate PassGAN, and the hyperparameters used in our
evaluation. We report on the evaluation of PassGAN, and on the comparison with
state-of-the-art password guessing techniques, in Section~\ref{sec:evaluation}.
We summarize our findings and discuss their implications, in
Section~\ref{sec:discussion}. We conclude in Section~\ref{sec:conclusion}.

\section{Background and Related Work}
\label{sec:background}

\subsection{Generative Adversarial Networks}
\label{sec:gan}

Generative Adversarial Networks (GANs) translate the current advances in deep neural networks for discriminative machine learning to (implicit) generative modeling. The goal of GANs is to generate samples from the same distribution as that of its training set $\mathcal{S}=\{\mathbf{x}_1, \mathbf{x}_2,\ldots, \mathbf{x}_n\}$. 
Generative modeling \cite{Murphy12} typically relies on closed-form expressions that, in many cases, cannot capture the nuisance of real data. GANs train a generative deep neural network \Generator{} that takes as input a multi-dimensional random sample $\mathbf{z}$ (from a Gaussian or uniform distribution) to generate a sample from the desired distribution. GANs transform the density estimation problem into a binary classification problem, in which the learning of the parameters of \Generator{} is achieved by relying on a discriminative deep neural network \Discriminator{} that needs to distinguish between the ``true'' samples in $\mathcal{S}$ and the ``fake'' samples produced by \Generator{}. More formally, the optimization problem solved by GANs can be summarized as follows:
\begin{equation}
\label{GAN}
\min_{\theta_G}\max_{\theta_D} \sum_{i=1}^{n} \log f(\mathbf{x}_i;\theta_D)+\sum_{j=1}^{n} \log(1-f(g(\mathbf{z}_j;\theta_G);\theta_D)),
\end{equation}
where $f(\mathbf{x};\theta_D)$ and $g(\mathbf{z}_j;\theta_G)$, respectively, represent \Discriminator{} and \Generator{}. The optimization shows the clash between the goals of the discriminator and generator deep neural networks. %
Since the original work by Goodfellow et al.~\cite{goodfellow2014generative}, there have been several improvements on GANs
~\cite{f-GAN, Bottou17, WesserstainGAN, Radford16, Goodfellow2016, Gulrajani2017ImprovedTO, MMD-GAN, Mroueh17, AdaGAN, Mescheder17-2, Kolter17, Roth17, Mroueh17-2, berthelot2017began, kim2017learning, mirza2014conditional, chen2016infogan, denton2015deep, zhang2016stackgan, Wei18,Petzka18,Hoang18,Cao18,Mroueh18,Binkowski18, Daskalakis18,  Hjelm18,  Miyato18, Zhou18}, where each new paper provides novel improvements in the domain. In this paper, we rely on IWGAN~\cite{Gulrajani2017ImprovedTO} as a building foundation for PassGAN, being that IWGAN~\cite{Gulrajani2017ImprovedTO} is among the first, most stable approaches for text generation via GANs. See Section~\ref{sec:iwgan_details} for more details on IWGAN. %

\subsection{Password Guessing}
\label{sec:passwordguessing}

Password guessing attacks are probably as old as password themselves~\cite{bidgoli2006handbook}, with more formal studies dating back to 1979~\cite{morris1979password}. In a password guessing attack, the adversary attempts to identify the password of one or more users by repeatedly testing multiple candidate passwords.

Two popular modern password guessing tools are John the Ripper (JTR)~\cite{jtr} and HashCat~\cite{hashcat}. Both tools implement multiple types of password guessing strategies, including: exhaustive brute-force attacks; dictionary-based attacks; rule-based attacks, which consist in generating password guesses from transformations of dictionary words~\cite{jtr_rules,hashcat_rules}; and Markov-model-based attacks~\cite{jtr_markov,hashcat_markov}.
JTR and HashCat are notably effective at guessing passwords. Specifically, there have been several instances in which well over 90\% of the passwords leaked from online services have been successfully recovered~\cite{pwproject}.

Markov models were first used to generate password guesses by Narayanan et al.~\cite{narayanan2005fast}. Their approach uses manually defined password rules, such as which portion of the generated passwords is composed of letters and numbers. Weir et al. ~\cite{weir2009password} subsequently improved this technique with Probabilistic Context-Free Grammars (PCFGs). With PCFGs, Weir et al.~\cite{weir2009password} demonstrated how to ``learn'' these rules from password distributions. Ma et al. ~\cite{ma2014study} and Durmuth et al.~\cite{durmuth2015omen} have subsequently extended this early work. %

To the best of our knowledge, the first work in the domain of passwords utilizing neural networks dates back to 2006 by Ciaramella et al.~\cite{ciaramella2006neural}. Recently, Melicher et al.~\cite{MelicherPassGuessNN} introduced \usenix, a
password guessing method based on recurrent neural networks~\cite{graves2013generating,sutskever2011generating}. 
However, the primary goal of these works consists in providing means for password strength estimation. For instance,  Melicher et al.~\cite{MelicherPassGuessNN}~aim at providing fast and accurate password strength estimation (thus \usenix~acronym),  while keeping the model as lightweight as possible, and minimizing accuracy loss. By keeping the model lightweight, \usenix~instantiates a password strength estimator that can be used in browsers through a (local) JavaScript implementation. To achieve this goal, \usenix~uses weight clipping without significantly sacrificing accuracy. In contrast, PassGAN focuses on the task of password guessing and attempts to do so with no a priori knowledge or assumption on the Markovian structure of user-chosen passwords.
%

\section{Experiment Setup}
\label{sec:sys_design}

To leverage the ability of GANs to estimate the probability effectively
distribution of passwords from the training set, we experimented with a variety
of parameters. In this section, we report our choices on specific GAN
architecture and hyperparameters.

\begin{figure}[t]
    \centering
        \includegraphics[width=\linewidth]{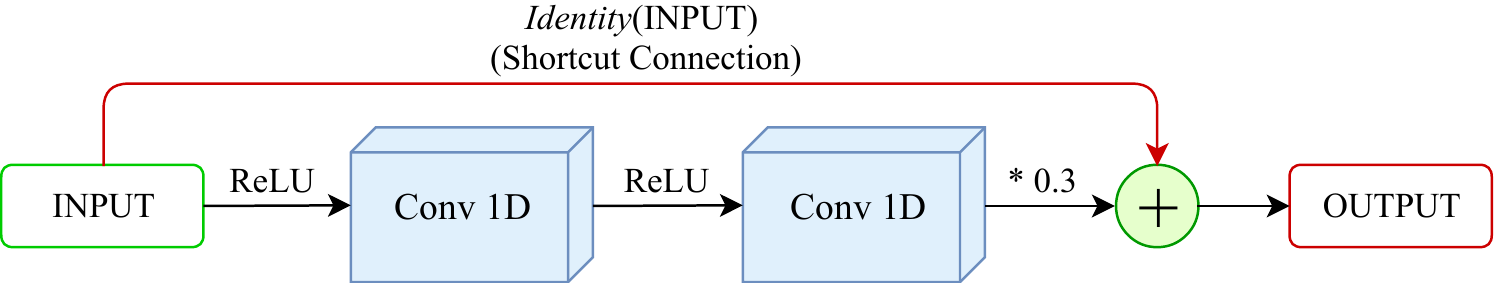}
    \caption{Representation of one Residual Block component constituting PassGAN}
    \label{fig:passgan_residual_block}
\end{figure}

We instantiated PassGAN using the {\em Improved training of Wasserstein GANs} (IWGAN) of Gulrajani et al.~\cite{Gulrajani2017ImprovedTO}. 
The IWGAN implementation used in this paper relies on the ADAM optimizer~\cite{kingma2014adam} to minimize the training error.%

\begin{figure*}
\centering
\begin{subfigure}{.85\textwidth}
  \centering
    \includegraphics[width=.95\linewidth]{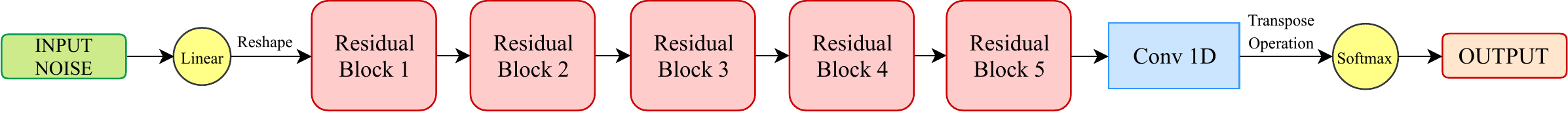}
  \caption{Generator Architecture, \Generator{}}
  \label{fig:passgan_generator_scheme}
\end{subfigure}
\begin{subfigure}{.85\textwidth}
  \centering
        \includegraphics[width=.95\linewidth]{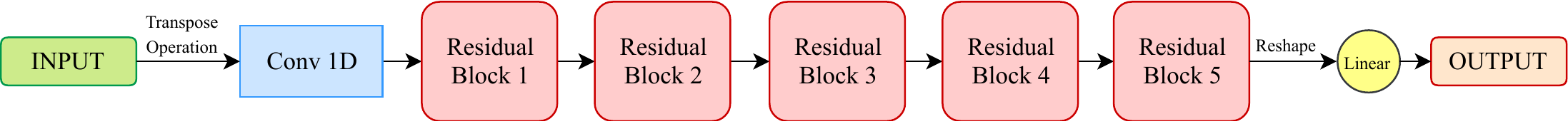}
  \caption{Discriminator Architecture, \Discriminator{}}
  \label{fig:passgan_discriminator_scheme}
\end{subfigure}
\caption{PassGAN's Architecture. During the training procedure, the discriminator \Discriminator{} processes passwords  from the training dataset, as well as password samples produced by the generator \Generator{}. Based on the feedback from \Discriminator{}, \Generator{} fine-tunes its parameters to produce password samples that are distributed similarly to the samples in the training set. In our setting, once the training procedure is complete, we use \Generator{} to generate password guesses.}
\label{fig:passgan_architecture}
\end{figure*}

The following hyper-parameters characterize our model: 
\begin{itemize}
\item {\bf Batch size}, which represents the number of passwords from the training set that propagate through the GAN at each step of the optimizer. We instantiated our model with a batch size of 64. %
\item {\bf Number of iterations}, which indicates how many times the GAN invokes its forward step and its back-propagation step~\cite{rumelhart1986learning, lecun1989backpropagation,lecun1990handwritten}. In each iteration, the GAN runs one generator iteration and one or more discriminator iterations. We trained the GAN using various number of iterations and eventually settled for 199,000 iterations, as further iterations provided diminishing returns in the number of matches. 

\item {\bf Number of discriminator iterations per generator iteration}, which indicates how many iterations the discriminator performs in each GAN iteration. The number of discriminator iterations per generative iteration was set to 10, which is the default value used by IWGAN.

\item {\bf Model dimensionality}, which represents the number of dimensions 
for each convolutional layer. We experimented using 5 residual layers for both the generator and the discriminator, with each of the layers in both deep neural networks having 128 dimensions.

\item {\bf Gradient penalty coefficient} ($\lambda$), %
which specifies the penalty applied to the norm of the gradient of the discriminator with respect to its input~\cite{Gulrajani2017ImprovedTO}. 
Increasing this parameter leads to a more stable training of the GAN~\cite{Gulrajani2017ImprovedTO}. In our experiments, we set the value of the gradient penalty to 10.%

\item {\bf Output sequence length}, which indicates the maximum length of the strings generated by the generator (\Generator{}). We modified the length of the sequence generated by the GAN from 32 characters (default length for IWGAN) to 10 characters, to match the maximum length of passwords used during training. %

\item {\bf Size of the input noise vector (seed)}, which determines how many random numbers from a normal distribution are fed as input to \Generator{} to generate samples. We set this size to 128 floating point numbers.

\item {\bf Maximum number of examples}, which represents the maximum number of training items (passwords, in the case of PassGAN) to load. The maximum number of examples loaded by the GAN was set to the size of the entire training dataset.%

\item {\bf Adam optimizer's hyper-parameters}:
    \begin{itemize}
    \item {\bf Learning rate}, i.e., how quickly the weights of the model are adjusted %
    \item {\bf Coefficient {\bf $\beta_1$}}, which specifies the decaying rate of the running average of the gradient.
    \item {\bf Coefficient {\bf $\beta_2$}}, which indicates the decaying rate of the running average of the square of the gradient. %
    \end{itemize}
    
    Coefficients $\beta_1$ and $\beta_2$ of the Adam optimizer were set to 0.5 and 0.9, respectively, while the learning rate was $10^{-4}$. These parameters are the default values used by Gulrajani et al.~\cite{Gulrajani2017ImprovedTO}.
\end{itemize}

Figure \ref{fig:passgan_residual_block} shows the structure of one residual block in PassGAN.
Figures \ref{fig:passgan_generator_scheme} and \ref{fig:passgan_discriminator_scheme} provide an overview of PassGAN's Generator and Discriminator models. In our experiments, we used 5 residual blocks in each model.

Our experiments were run using the TensorFlow implementation of IWGAN found at~\cite{iwgan_code}. %
We used TensorFlow version 1.2.1 for GPUs~\cite{abadi2016tensorflow}, with Python version 2.7.12. All experiments were performed on a workstation running Ubuntu 16.04.2 LTS, with 64GB of RAM, a 12-core 2.0 GHz Intel Xeon CPU, and an NVIDIA GeForce GTX 1080 Ti GPU with 11GB of global memory.

\paragraph{IWGAN}
\label{sec:iwgan_details}

The building blocks used to construct IWGAN~\cite{Gulrajani2017ImprovedTO}, and consequently PassGAN, are Residual Blocks, an example of which is shown in Figure \ref{fig:passgan_residual_block}. They are the central component of Residual Networks (ResNets), introduced by He et al., \cite{He2016DeepRL} at CVPR'16. 
When training a deep neural network, initially the training error decreases as the number of layer increases. However, after reaching a certain number of layers, training error starts increasing again. %

To address this problem, He et al.~\cite{He2016DeepRL} introduced ResNet. In contrast with other deep neural networks, ResNet includes ``shortcut connection'' between layers \cite{resnetReview}. This can be seen as a wrapper for these layers and is implemented as the identity function (indicated as Residual Block in Figure~\ref{fig:passgan_residual_block}). By using multiple consecutive residual blocks, ResNet consistently reduces training error as the number of layers increases. %

Residual Blocks in PassGAN are composed of two 1-dimensional  convolutional layers, connected with one another with rectified linear units (ReLU) activation functions, Figure \ref{fig:passgan_residual_block}. The input of the block is the identity function, and is increased with $0.3 \cdot $\emph{output of convolutional layers} to produce the output of the block. Figures \ref{fig:passgan_generator_scheme} and \ref{fig:passgan_discriminator_scheme} provide a schematic view of PassGAN's Generator and Discriminator models. In our experiments, we used 5 residual blocks, (see Figure~\ref{fig:passgan_residual_block}).

\subsection{Training and Testing}
\label{sec:gan_traininig_testing}

To evaluate the performance of PassGAN, and to compare it with state-of-the-art
password generation rules, we first trained the GAN, JTR, HashCat,
the Markov model, PCFG, and \usenix~on a large set of passwords from the RockYou
password leak~\cite{rockyou_dataset}.\footnote{We consider the use of publicly
available password datasets to be ethical, and consistent with security research
best practices (see, e.g., \cite{dell2010password, MelicherPassGuessNN,
castelluccia2012adaptive}).} Entries in this dataset represent a mixture of
common and complex passwords.%
\paragraph{RockYou Dataset}
The RockYou dataset~\cite{rockyou_dataset} contains 32,503,388 passwords. We
selected all passwords of length 10 characters or less (29,599,680 passwords,
which correspond to 90.8\% of the dataset), and used 80\% of them (23,679,744
total passwords, 9,926,278 unique passwords) to train each password guessing
tool. We refer the reader to Section~\ref{sec:password_sampling_other_tools} for further details on the training procedure of each tool. For testing, we computed the (set) difference between the remaining 20\% of the
dataset (5,919,936 total passwords, 3,094,199 unique passwords) and the training
test. The resulting 1,978,367 entries correspond to passwords that were not
previously observed by the password guessing tools. This allowed us to count
only non-trivial matches in the testing set. %

\paragraph{LinkedIn Dataset}
We also tested each tool on passwords from the LinkedIn dataset~\cite{linkedin_dataset}, of length up to 10 characters, and that were not present in the training set. The LinkedIn dataset consists of 60,065,486 total unique passwords (43,354,871 unique passwords with length 10 characters or less), out of which 40,593,536 were not in the training dataset from RockYou. (Frequency counts were not available for the LinkedIn dataset.) Passwords in the LinkedIn dataset were exfiltrated as hashes, rather than in plaintext. As such, the LinkedIn dataset contains only plaintext passwords that tools such as JTR and HashCat were able to recover, thus giving rule-based systems a potential edge.

Our training and testing procedures showed: %
\begin{inparaenum}[(1)]
\item how well PassGAN predicts passwords when trained and tested on the same password distribution (i.e., when using the RockYou dataset for both training and testing); and
\item whether PassGAN generalizes across password datasets, i.e., how it performs when trained on the RockYou dataset, and tested on the LinkedIn dataset.
\end{inparaenum}

\subsection{Password Sampling Procedure for HashCat, JTR, Markov Model, PCFG and \usenix}
\label{sec:password_sampling_other_tools}
We used the portion of RockYou dataset selected for training, see Section~\ref{sec:gan_traininig_testing}, as the input dataset to HashCat Best64, HashCat gen2, JTR Spiderlab rules, Markov Model, PCFG, and \usenix, and generated passwords as follows:

\begin{itemize}

\item We instantiated HashCat and JTR's rules using passwords from the training set sorted by frequency in descending order (as in~\cite{MelicherPassGuessNN}). HashCat Best64 generated 754,315,842 passwords, out of which 361,728,683 were unique and of length 10 characters or less. Note that this was the maximum number of samples produced by Best64 rule-set for the given input set, i.e., RockYou training set. With HashCat gen2 and JTR SpiderLab we uniformly sampled a random subset of size $10^9$ from their output. This subset was composed of passwords of length 10 characters or less.

\item For \usenix, we set up the code from~\cite{CMUsourcecode} according to the
instruction provided in~\cite{CMUinstallationinstructions}. We trained a model
containing 2-hidden layers and 1 dense layer of size 512 (for the full list of parameters see Table~\ref{tab:cmu_parameters} in Appendix~\ref{sec:cmu_parameters}). 
We did not perform any transformation (e.g., removing symbols, or transforming
all characters to lowercase) on the training set for the sake of consistency
with the other tools. Once trained, \usenix~enumerates a subset of its output
space defined by a probability threshold $p$: a password belongs to \usenix's
output if and only if its estimated probability is at least $p$. In our experiments, we set
$p=10^{-10}$. This resulted in a total of 747,542,984 passwords of length 10
characters or less. Before using these passwords in our evaluation, we sorted
them by probability in descending order.

\item We generated 494,369,794 unique passwords of length 10 or less using
the 3-gram Markov model. We ran this model using its standard
configuration~\cite{dorseyMarkovCode}.

\item We generated $10^{9}$ unique passwords of length 10 or less using the PCFG implementation of Weir et al.~\cite{weirpcfgimplementation09}.
\end{itemize}

\section{Evaluation}
\label{sec:evaluation}

In this section, we present our evaluation steps. We first evaluate the number
of matches generated by PassGAN's output, and then compare it with \usenix, with
a popular 3-gram implementation of Markov models~\cite{dorseyMarkovCode}, with PCFGs~\cite{weirpcfgimplementation09}, and
with password generation rules for JTR (SpiderLab mangling
rules~\cite{spiderlabs}) and HashCat (Best64 and gen2 rules~\cite{hashcat}).
These password generation rules are commonly used in the password guessing
literature (see, e.g.,~\cite{MelicherPassGuessNN}), and have been optimized over
several years on password datasets including RockYou and LinkedIn. Because of
these dataset-specific optimizations, we consider these rules a good
representation of the best matching performance that can be obtained with
rules-based password guessing. Further, we provide experimental results
evaluating PassGAN in combination with HashCat Best64. We conclude the section
providing a comparison between PassGAN and \usenix~in terms of probability
densities and password distribution.

\begin{table}[]
    \caption{Number of passwords generated by PassGAN that match passwords in the  RockYou testing set. Results are shown in terms of unique matches.}%
    \label{tab:passgan_matches}
    \footnotesize
    \begin{tabular}{|c|c|c|}
    \hline
    \textbf{\begin{tabular}[c]{@{}c@{}}Passwords\\ Generated\end{tabular}} & 
    \textbf{\begin{tabular}[c]{@{}c@{}}Unique\\ Passwords\end{tabular}} & 
    \textbf{\begin{tabular}[c]{@{}c@{}}Passwords matched in testing\\ set, and not in training set\\ (1,978,367 unique samples)\end{tabular}} \\ \hline
    $10^{4}$ & 9,738 & 103 (0.005\%) \\ \hline
    $10^{5}$ &94,400 & 957 (0.048\%) \\ \hline
    $10^{6}$ &855,972 & 7,543 (0.381\%) \\ \hline
    $10^{7}$ &7,064,483 & 40,320 (2.038\%) \\ \hline
    $10^{8}$ &52,815,412 & 133,061 (6.726\%) \\ \hline
    $10^{9}$ &356,216,832 & 298,608 (15.094\%) \\ \hline
    $10^{10}$ &2,152,819,961 & 515,079 (26.036\%) \\ \hline
     \hline
    $2 \cdot 10^{10}$ &3,617,982,306 & 584,466 (29.543\%) \\ \hline
    $3 \cdot 10^{10}$ &4,877,585,915 & 625,245 (31.604\%) \\ \hline
    $4 \cdot 10^{10}$ &6,015,716,395 & 653,978 (33.056\%) \\ \hline
    $5 \cdot 10^{10}$ &7,069,285,569 & 676,439 (34.192\%) \\ \hline
    \end{tabular}
\end{table}

\subsection{PassGAN's Output Space}

To evaluate the size of the password space generated by PassGAN, we generated several password sets of sizes between $10^{4}$ and $10^{10}$. Our experiments show that, as the number of passwords increased, so did the number of unique (and therefore new) passwords generated. Results of this evaluation are reported in Table~\ref{tab:passgan_matches}.

When we increased the number of passwords generated by PassGAN, the rate at which new unique passwords were generated decreased only slightly. Similarly, the rate of increase of the number of matches (shown in Table~\ref{tab:passgan_matches}) diminished slightly as the number of passwords generated increased. This is to be expected, as the simpler passwords are matched early on, and the remaining (more complex) passwords require a substantially larger number of attempts in order to be matched.

\begin{figure}[h]
\centering
\includegraphics[width=3.1in]{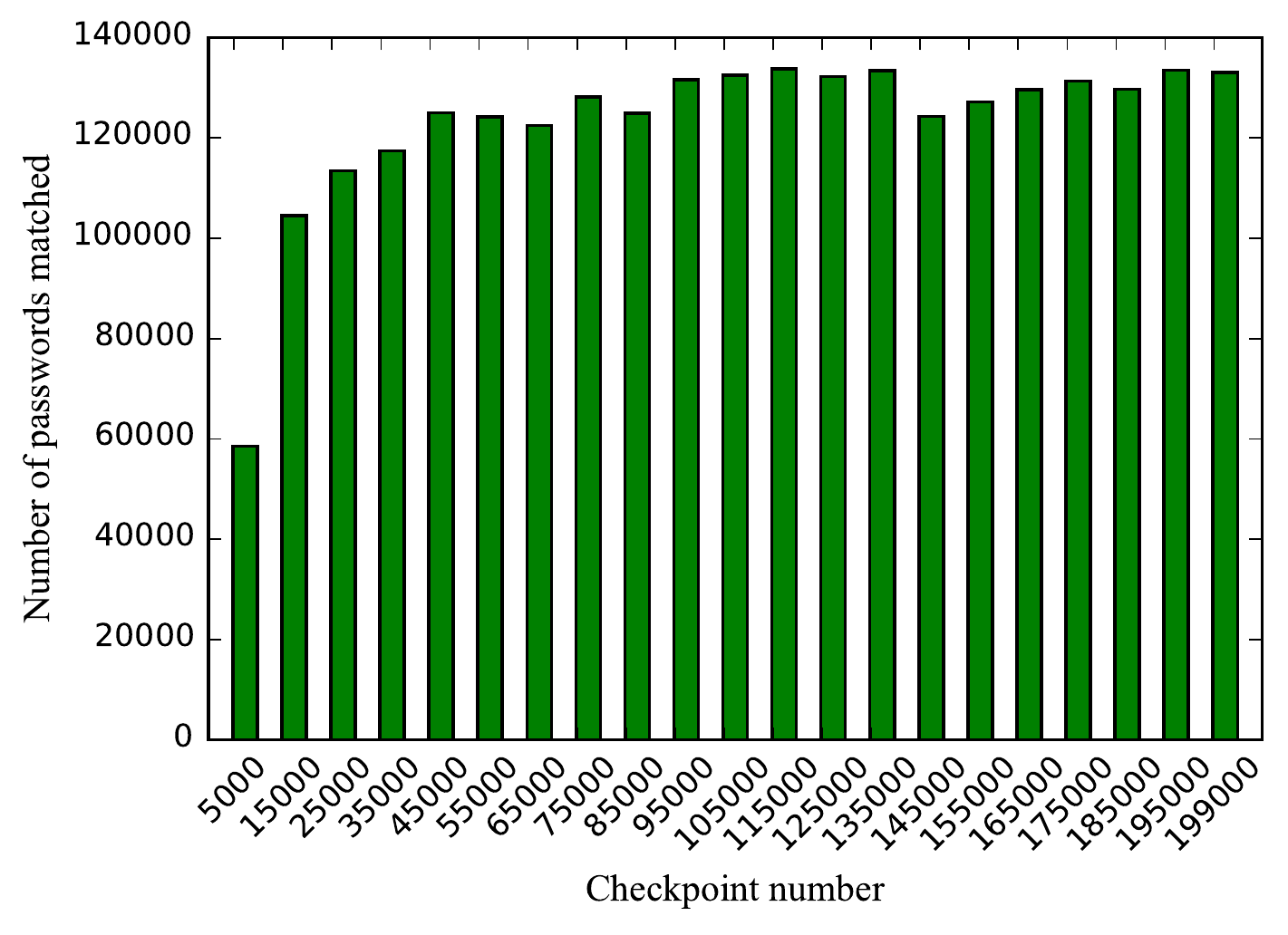}
\caption{Number of unique passwords generated by PassGAN on various checkpoints, matching the RockYou testing set. The $x$ axis represents the number of  iterations (checkpoints) of PassGAN's training process. For each checkpoint, we sampled $10^{8}$ passwords from PassGAN.
}
\label{fig:training_progress}
\end{figure}%

\paragraph{Impact of Training Process on Overfitting.} 
\label{sec:gan_iterations}
Training a GAN is an iterative process that consists of a large number of 
iterations.   
As the number of iterations increases, the GAN learns more information from the
distribution of the data. However, increasing the number of steps also increases
the probability of
overfitting~\cite{goodfellow2014generative, wu2016quantitative}.

To evaluate this tradeoff on password data, we stored intermediate training
checkpoints and generated $10^8$ passwords at each checkpoint. Figure
\ref{fig:training_progress} shows how many of these passwords match with the
content of the RockYou testing set. 
In general, the number of matches increases with the number of iterations. This
increase tapers off around 125,000-135,000 iterations, and then again around
190,000-195,000 iterations, where we stopped training the GAN. This indicates
that further increasing the number of iterations will likely lead to
overfitting, thus reducing the ability of the GAN to generate a wide variety of
highly likely passwords. Therefore, we consider this range of iterations
adequate for the RockYou training~set. %

\begin{table*}[]
\centering
\footnotesize
\caption{Number of matches generated by each password guessing tool against the RockYou testing set, and corresponding number of password generated by PassGAN to outperform each tool. Matches for HashCat Best64 and \usenix~were obtained by exhaustively enumerating the entire output of each tool. The minimum probability threshold for \usenix~was set to $p=10^{-10}$.}
\label{tab:passgan_outperfom}
\begin{tabular}{|c|c|c|c||c|}
\hline
\textbf{Approach} & \textbf{\begin{tabular}[c]{@{}c@{}}(1) Unique\\ Passwords\end{tabular}} & \textbf{(2) Matches} & \textbf{\begin{tabular}[c]{@{}c@{}}(3) Number of passwords \\ required for PassGAN \\ to outperform (2)\end{tabular}} &  \textbf{\begin{tabular}[c]{@{}c@{}}(4) PassGAN\\ Matches\end{tabular}} \\ \hline
\textbf{\begin{tabular}[c]{@{}c@{}}JTR\\ Spyderlab\end{tabular}} & $10^{9}$ & 461,395 (23.32\%) & $1.4 \cdot 10^{9}$ & 461,398 (23.32\%) \\ \hline 
\textbf{\begin{tabular}[c]{@{}c@{}}Markov Model\\ 3-gram\end{tabular}} & $4.9 \cdot 10^{8} $ & 532,961 (26.93\%) &  $2.47 \cdot 10^9$ & 532,962 (26.93\%) \\ \hline 
\textbf{\begin{tabular}[c]{@{}c@{}}HashCat\\ gen2\end{tabular}} & $10^{9}$ & 597,899 (30.22\%) & $4.8 \cdot 10^{9}$ & 625,245 (31.60\%) \\ \hline
\textbf{\begin{tabular}[c]{@{}c@{}}HashCat\\ Best64\end{tabular}} & $3.6 \cdot 10^{8}$ & 630,068 (31.84\%) & $5.06 \cdot 10^{9}$ & 630,335 (31.86\%) \\ \hline
\textbf{\begin{tabular}[c]{@{}c@{}}PCFG\end{tabular}} & $10^{9}$ & 486,416 (24.59\%) & $2.1 \cdot 10^{9}$ & 511,453 (25.85\%) \\ \hline
\textbf{\begin{tabular}[c]{@{}c@{}}\usenix{}\\ $p=10^{-10}$\end{tabular}} & $7.4 \cdot 10^{8}$ & 652,585 (32.99\%) & $6 \cdot 10^{9}$ & 653,978 (33.06\%) \\ \hline
\end{tabular}
\end{table*}

\subsection{Evaluating the Passwords Generated by PassGAN}
To evaluate the quality of the output of PassGAN, we generated $5\cdot10^{10}$ passwords, out of which roughly $7\cdot10^9$ were unique. We compared these passwords with the outputs of length 10 characters or less from HashCat Best64, HashCat gen2, JTR SpiderLab, \usenix, PCFG, and Markov model, see Section~\ref{sec:password_sampling_other_tools} for the configuration and sampling procedures followed for each of these tools.

In our comparisons, we aimed at establishing whether PassGAN was able to meet the
performance of the other tools, \emph{despite its lack of any a-priori knowledge on
password structures}. This is because we are primarily interested in determining
whether the properties that PassGAN autonomously extracts from a list of
passwords can represent enough information to compete with state-of-the-art
human-generated rules and Markovian password generation processes.

Our results show that, for each of the tools, PassGAN was able to generate at
least the same number of matches. Additionally, to achieve this result, PassGAN
needed to generate a number of passwords that was within one order of magnitude
of each of the other tools. This holds for both the RockYou and the 
LinkedIn testing sets. This is not unexpected, because while other tools rely on prior knowledge on passwords for guessing, PassGAN does not. 
Table~\ref{tab:passgan_outperfom} summarizes our findings for the RockYou
testing set, while Table~\ref{tab:passgan_outperfom_linkedin} shows our results
for the LinkedIn test set.

Our results also show that PassGAN has an advantage with respect to rule-based
password matching when guessing passwords from a dataset different from the one
it was trained on. In particular, PassGAN was able to match more passwords than
HashCat within a smaller number of attempts ($2.1 \cdot 10^9$ -- $3.6 \cdot 10^9$ for LinkedIn, compared to $4.8\cdot 10^9$ -- $5.06 \cdot10^9$ for RockYou).

\begin{table*}[]
\centering
\footnotesize
\caption{Number of matches generated by each password guessing tool against the LinkedIn testing set, and corresponding number of password generated by PassGAN to outperform each tool. Matches for HashCat Best64 and \usenix~were obtained by exhaustively enumerating the entire output of each tool. The minimum probability threshold for \usenix~was set to $p=10^{-10}$.} 
\label{tab:passgan_outperfom_linkedin}
\begin{tabular}{|c|c|c|c||c|}
\hline
\textbf{Approach} & \textbf{\begin{tabular}[c]{@{}c@{}}(1) Unique\\ Passwords\end{tabular}} & \textbf{(2) Matches} & \textbf{\begin{tabular}[c]{@{}c@{}}(3) Number of passwords \\ required for PassGAN \\ to outperform (2)\end{tabular}} &  \textbf{\begin{tabular}[c]{@{}c@{}}(4) PassGAN\\ Matches\end{tabular}} \\ \hline
\textbf{\begin{tabular}[c]{@{}c@{}}JTR\\ Spyderlab\end{tabular}} & $10^{9}$ & 6,840,797 (16.85\%) & $2.7 \cdot 10^9$ & 6,841,217 (16.85\%) \\ \hline  
\textbf{\begin{tabular}[c]{@{}c@{}}Markov Model\\ 3-gram\end{tabular}} & $4.9 \cdot 10^{8} $ & 5,829,786 (14.36\%) & $1.6 \cdot 10^{9}$ &  5,829,916 (14.36\%) \\ \hline 
\textbf{\begin{tabular}[c]{@{}c@{}}HashCat\\ gen2\end{tabular}} & $10^{9}$ & 6,308,515 (15.54\%) & $2.1\cdot 10^9$ & 6,309,799 (15.54\%) \\ \hline 
\textbf{\begin{tabular}[c]{@{}c@{}}HashCat\\ Best64\end{tabular}} & $3.6 \cdot 10^{8}$ & 7,174,990 (17.67\%) & $3.6 \cdot 10^{9}$ & 7,419,248 (18.27\%) \\ \hline
\textbf{\begin{tabular}[c]{@{}c@{}}PCFG\end{tabular}} & $10^{9}$ & 7,288,553 (17.95\%) & $3.6 \cdot 10^{9}$ & 7,419,248 (18.27\%) \\ \hline
\textbf{\begin{tabular}[c]{@{}c@{}}\usenix{}\\ $p=10^{-10}$\end{tabular}} & $7.4 \cdot 10^{8}$ & 8,290,173 (20.42\%) & $6 \cdot 10^{9}$ & 8,519,060 (21.00\%) \\ \hline
\end{tabular}
\end{table*}

\subsection{Combining PassGAN with HashCat}
To maximize the number of passwords guessed, the adversary would typically use the output of multiple tools in order to combine the benefits of rule-based tools (e.g., fast password generation) and ML-based tools (e.g., generation of a large number of guesses). %

To evaluate PassGAN in this setting, we removed all passwords matched by HashCat Best64 (the best performing set of rules in our experiments) from the RockYou and LinkedIn testing sets. This led to two new test sets, containing 1,348,300 (RockYou) and 33,394,178 (LinkedIn) passwords, respectively. 

Our results show that the number of matches steadily increases with the number
of samples produced by PassGAN. In particular, when we used $7\cdot10^9$
passwords from PassGAN, we were able to match 51\% (320,365) of passwords from
the ``new'' RockYou dataset, and 73\% (5,262,427) additional passwords from the
``new'' LinkedIn dataset. This confirms that combining rules
with machine learning password guessing is an effective strategy. 
Moreover, it confirms that PassGAN can
capture portions of the password space not covered by rule-based approaches.
With this in mind, a recent version of HashCat~\cite{hashcatv5} introduced a
generic password candidate interface called ``slow candidates'', enabling the
use of tools such as PCFGs~\cite{weir2009password}, OMEN~\cite{durmuth2015omen}, PassGAN, and more with HashCat.

\subsection{Comparing PassGAN with \usenix}

In this section, we concentrate on comparing PassGAN with \usenix~having a particular focus on the probability estimation. %
\usenix~is based on recurrent neural networks~\cite{graves2013generating, sutskever2011generating}, and typically the model is trained on password leaks from several websites, in our case the RockYou training set.
During password generation, the neural network generates one password character at a time. Each new character (including a special end-of-password character) is computed based on its probability, given the current output state, in what is essentially a Markov process. 
Given a trained \usenix~model, FLA outputs the following six fields:
\begin{inparaenum}
    \item password,
    \item the probability of that password,
    \item the estimated output guess number, i.e., the strength of that password,
    \item the standard deviation of the randomized trial for this password (in units of the number of guesses),
    \item the number of measurements for this password and
    \item the estimated confidence interval for the guess number (in units of the number of guesses).
\end{inparaenum}
The evaluation presented in \cite{MelicherPassGuessNN} shows that their technique outperforms Markov models, PCFGs and password composition rules commonly used with JTR and HashCat, when testing a large number of password guesses (in the $10^{10}$ to $10^{25}$ range).

We believe that one of the limitations of \usenix~resides precisely in the Markovian nature of the process used to estimate passwords. For instance, {\tt 123456}; {\tt 12345}; and, {\tt 123456789} are the three most common passwords in the RockYou dataset, being roughly one every 66-passwords. Similarly, the most common passwords produced by \usenix~start with ``{\tt 123}'' or use the word ``{\tt love}'', Table \ref{tab:FLA_density}. In contrast, PassGAN's most commonly generated passwords, Table~\ref{tab:passgan_density}, tend to show more variability with samples composed of names, the combination of names and numbers, and more. When compared with Table \ref{tab:ry_training}, the most likely samples from PassGAN exhibit closer resemblance to the training set and its probabilities than FLA does. 
We argue that due to the Markovian structure of the password generation process in \usenix,~any password characteristic that is not captured within the scope of an $n-$gram, might not be encoded by \usenix. For instance, if a meaningful subset of 10-character passwords is constructed as the concatenation of two words (e.g., {\tt MusicMusic}), any Markov process with $n \leq 5$ will not be able to capture this behavior properly. On the other hand, given enough examples, the neural network used in PassGAN will be able to learn this property. As a result, while password {\tt pookypooky} was assigned a probability $p\approx10^{-33}$ by \usenix~(with an estimated number of guessing attempts of about $10^{29}$), it was guessed after roughly $10^8$ attempts by PassGAN. 

\begin{figure}[]
\centering
\includegraphics[width=\columnwidth]{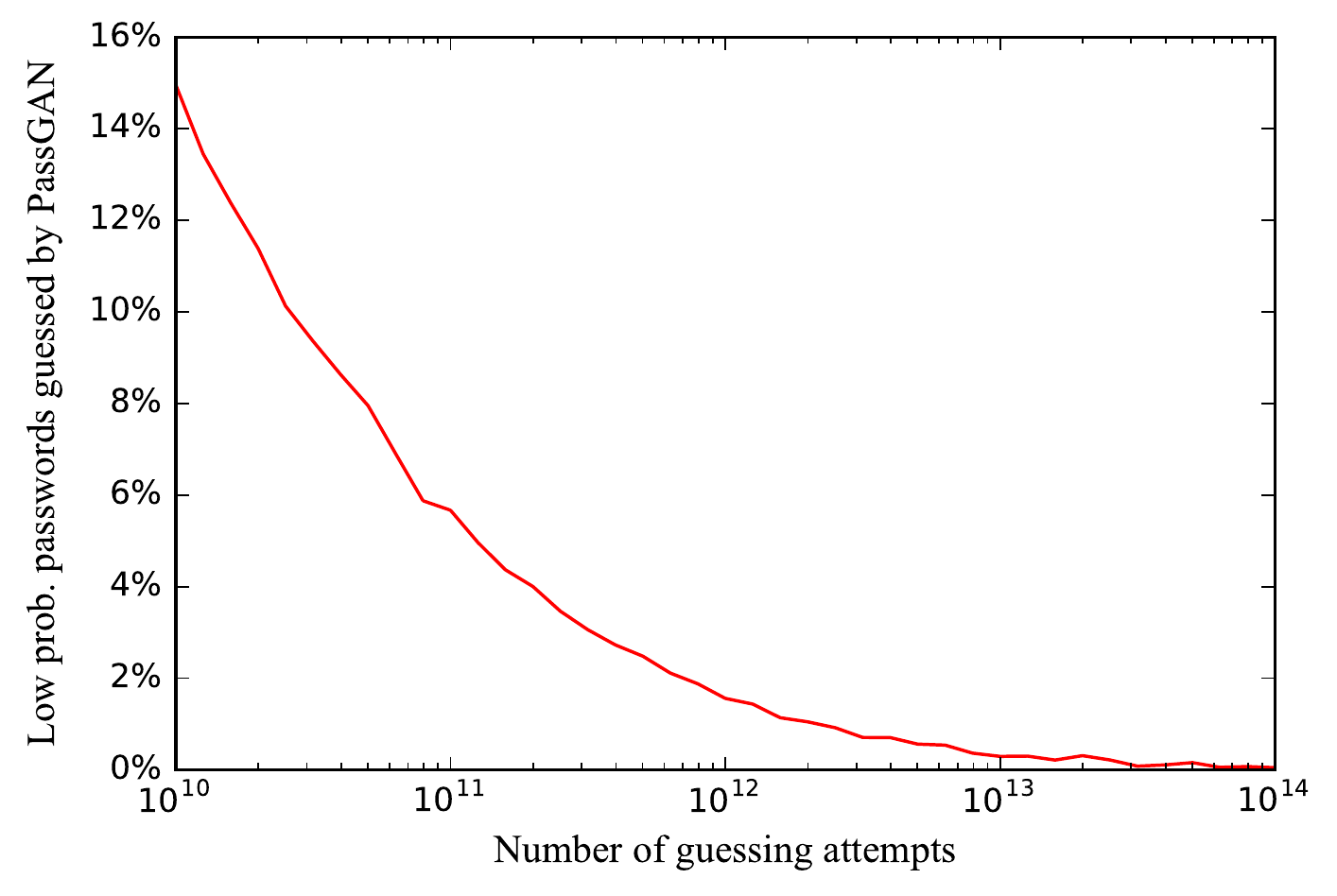}
\caption{Percentage of passwords matched by \usenix~at a particular number of guesses, that are matched by PassGAN in at most $7\cdot 10^{9}$ attempts.}
\label{fig:passgan_cmu_strong_guesses}
\end{figure}

To investigate further on the differences between PassGAN and \usenix, we computed the number of passwords in the RockYou testing set for which \usenix~required at least $10^{10}$ attempts and that PassGAN was able to guess within its first $7\cdot10^9$ samples. These are the passwords to which \usenix~assigns low probabilities, despite being chosen by some users. Because PassGAN can model them, we conclude that the probabilities assigned by \usenix~to these passwords are incorrect. Figure~\ref{fig:passgan_cmu_strong_guesses} presents our result as the ratio between the passwords matched by \usenix~at a
particular number of guessing attempts, and by PassGAN within its first $7\cdot10^9$ attempts. Our results show that PassGAN can model a number of passwords more correctly than \usenix. However, this advantage decreased as the number of attempts required for \usenix~to guess a password increased, i.e., as the estimated probability of that password decreased. This shows that, in general, the two tools agree on assigning probabilities to passwords. 

\begin{table*}[]
    \caption{(a) Top-50 passwords present in RockYou training dataset sorted by frequency.
(b) Frequency of the 50 most common outputs of \usenix, and corresponding
frequency and rank in the RockYou training set. Passwords are sorted by the
probability assigned by \usenix. (c) Frequency of the 50 most common outputs of PassGAN, and corresponding
frequency and rank in the RockYou training set. Passwords are sorted by the
frequency in which they appear in PassGAN's output. ``---'' indicates that the password was not in the training set.}
\label{tab:passgan_fla_allinone}
\begin{subtable}{.3\linewidth}
\centering
\scriptsize
\caption{RockYou Training Set}
\label{tab:ry_training}
\begin{tabular}{c|c|c}
\textbf{Password}  & \textbf{\begin{tabular}[c]{@{}c@{}}Number of \\ Occurrences in\\Training Set\end{tabular}} & \textbf{\begin{tabular}[c]{@{}c@{}}Frequency in\\ Training Set\end{tabular}} \\
\hline
\tt     123456 & 232,844 & 0.9833\%  \\ 
\tt     12345 & 63,135 & 0.2666\%  \\ 
\tt     123456789 & 61,531 & 0.2598\%  \\ 
\tt     password & 47,507 & 0.2006\%  \\ 
\tt     iloveyou & 40,037 & 0.1691\%  \\ 
\tt     princess & 26,669 & 0.1126\%  \\ 
\tt     1234567 & 17,399 & 0.0735\%  \\ 
\tt     rockyou & 16,765 & 0.0708\%  \\ 
\tt     12345678 & 16,536 & 0.0698\%  \\ 
\tt     abc123 & 13,243 & 0.0559\%  \\ 
\tt     nicole & 12,992 & 0.0549\%  \\ 
\tt     daniel & 12,337 & 0.0521\%  \\ 
\tt     babygirl & 12,130 & 0.0512\%  \\ 
\tt     monkey & 11,726 & 0.0495\%  \\ 
\tt     lovely & 11,533 & 0.0487\%  \\ 
\tt     jessica & 11,262 & 0.0476\%  \\ 
\tt     654321 & 11,181 & 0.0472\%  \\ 
\tt     michael & 11,174 & 0.0472\%  \\ 
\tt     ashley & 10,741 & 0.0454\%  \\ 
\tt     qwerty & 10,730 & 0.0453\%  \\ 
\tt     iloveu & 10,587 & 0.0447\%  \\ 
\tt     111111 & 10,529 & 0.0445\%  \\ 
\tt     000000 & 10,412 & 0.0440\%  \\ 
\tt     michelle & 10,210 & 0.0431\%  \\ 
\tt     tigger & 9,381 & 0.0396\%  \\ 
\tt     sunshine & 9,252 & 0.0391\%  \\ 
\tt     chocolate & 9,012 & 0.0381\%  \\ 
\tt     password1 & 8,916 & 0.0377\%  \\ 
\tt     soccer & 8,752 & 0.0370\%  \\ 
\tt     anthony & 8,564 & 0.0362\%  \\ 
\tt     friends & 8,557 & 0.0361\%  \\ 
\tt     butterfly & 8,427 & 0.0356\%  \\ 
\tt     angel & 8,425 & 0.0356\%  \\ 
\tt     purple & 8,381 & 0.0354\%  \\ 
\tt     jordan & 8,123 & 0.0343\%  \\ 
\tt     liverpool & 7,846 & 0.0331\%  \\ 
\tt     loveme & 7,818 & 0.0330\%  \\ 
\tt     justin & 7,769 & 0.0328\%  \\ 
\tt     fuckyou & 7,702 & 0.0325\%  \\ 
\tt     football & 7,559 & 0.0319\%  \\ 
\tt     123123 & 7,545 & 0.0319\%  \\ 
\tt     secret & 7,458 & 0.0315\%  \\ 
\tt     andrea & 7,395 & 0.0312\%  \\ 
\tt     carlos & 7,281 & 0.0307\%  \\ 
\tt     jennifer & 7,229 & 0.0305\%  \\ 
\tt     joshua & 7,186 & 0.0303\%  \\ 
\tt     bubbles & 7,031 & 0.0297\%  \\ 
\tt     1234567890 & 6,953 & 0.0294\%  \\ 
\tt     hannah & 6,911 & 0.0292\%  \\ 
\tt     superman & 6,855 & 0.0289\%  \\ 
\hline
\end{tabular}
\end{subtable}%
\begin{subtable}{.35\linewidth}
      \centering
        \caption{\usenix}
        \label{tab:FLA_density}
        \scriptsize
\begin{tabular}{c|c|c|c}
\textbf{Password}  & \textbf{\begin{tabular}[c]{@{}c@{}}Rank in\\ Training Set\end{tabular}} & \textbf{\begin{tabular}[c]{@{}c@{}}Frequency in\\ Training Set\end{tabular}} &\textbf{\begin{tabular}[c]{@{}c@{}}Probability \\assigned by \\ \usenix\end{tabular}} \\
\hline
\tt     123456 & 1 & 0.9833\% & 2.81E-3 \\ 
\tt     12345 & 2 & 0.2666\% & 1.06E-3 \\ 
\tt     123457 & 3,224 & 0.0016\% & 2.87E-4 \\ 
\tt     1234566 & 5,769 & 0.0010\% & 1.85E-4 \\ 
\tt     1234565 & 9,692 & 0.0006\% & 1.11E-4 \\ 
\tt     1234567 & 7 & 0.0735\% & 1.00E-4 \\ 
\tt     12345669 & 848,078 & 0.0000\% & 9.84E-5 \\ 
\tt     123458 & 7,359 & 0.0008\% & 9.54E-5 \\ 
\tt     12345679 & 7,818 & 0.0007\% & 9.07E-5 \\ 
\tt     123459 & 8,155 & 0.0007\% & 7.33E-5 \\ 
\tt     lover & 457 & 0.0079\% & 6.73E-5 \\ 
\tt     love & 384 & 0.0089\% & 6.09E-5 \\ 
\tt     223456 & 69,163 & 0.0001\% & 5.14E-5 \\ 
\tt     22345 & 118,098 & 0.0001\% & 4.61E-5 \\ 
\tt     1234564 & 293,340 & 0.0000\% & 3.81E-5 \\ 
\tt     123454 & 23,725 & 0.0003\% & 3.56E-5 \\ 
\tt     1234569 & 5,305 & 0.0010\% & 3.54E-5 \\ 
\tt     lovin & 39,712 & 0.0002\% & 3.21E-5 \\ 
\tt     loven & 57,862 & 0.0001\% & 3.09E-5 \\ 
\tt     iloveyou & 5 & 0.1691\% & 3.05E-5 \\ 
\tt     1234568 & 6,083 & 0.0009\% & 2.99E-5 \\ 
\tt     223455 & 1,699,287 & 0.0000\% & 2.68E-5 \\ 
\tt     12345668 & 1,741,520 & 0.0000\% & 2.64E-5 \\ 
\tt     1234561 & 12,143 & 0.0005\% & 2.59E-5 \\ 
\tt     123455 & 5,402 & 0.0010\% & 2.58E-5 \\ 
\tt     ilover & 45,951 & 0.0002\% & 2.52E-5 \\ 
\tt     love15 & 2,074 & 0.0025\% & 2.39E-5 \\ 
\tt     12345660 &  ---  &  --- &2.39E-5 \\ 
\tt     1234560 & 3,477 & 0.0015\% & 2.34E-5 \\ 
\tt     123456789 & 3 & 0.2598\% & 2.27E-5 \\ 
\tt     love11 & 1,735 & 0.0029\% & 2.21E-5 \\ 
\tt     12345667 & 252,961 & 0.0000\% & 2.11E-5 \\ 
\tt     12345678 & 9 & 0.0698\% & 2.11E-5 \\ 
\tt     223457 & 8,929,184 & 0.0000\% & 2.09E-5 \\ 
\tt     love12 & 565 & 0.0067\% & 2.05E-5 \\ 
\tt     lovo &  ---  &  --- &2.03E-5 \\ 
\tt     12345666 & 46,540 & 0.0002\% & 1.98E-5 \\ 
\tt     123456689 &  ---  &  --- &1.97E-5 \\ 
\tt     1234562 & 92,917 & 0.0001\% & 1.92E-5 \\ 
\tt     12345699 & 197,906 & 0.0000\% & 1.90E-5 \\ 
\tt     123451 & 9,950 & 0.0006\% & 1.89E-5 \\ 
\tt     123450 & 7,186 & 0.0008\% & 1.88E-5 \\ 
\tt     loves & 779 & 0.0054\% & 1.83E-5 \\ 
\tt     1234576 & 61,296 & 0.0001\% & 1.80E-5 \\ 
\tt     love13 & 1,251 & 0.0038\% & 1.78E-5 \\ 
\tt     lovele & 154,468 & 0.0001\% & 1.78E-5 \\ 
\tt     lovine & 4,497,922 & 0.0000\% & 1.75E-5 \\ 
\tt     lovi & 4,498,263 & 0.0000\% & 1.74E-5 \\ 
\tt     iloven & 323,339 & 0.0000\% & 1.59E-5 \\ 
\tt     lovina & 62,446 & 0.0001\% & 1.53E-5 \\ 
\hline
\end{tabular}
    \end{subtable}%
 \begin{subtable}{.35\linewidth}
      \centering
        \caption{PassGAN}
        \label{tab:passgan_density}
        \scriptsize
\begin{tabular}{c|c|c|c}
\textbf{Password}  & \textbf{\begin{tabular}[c]{@{}c@{}}Rank in\\ Training Set\end{tabular}} & \textbf{\begin{tabular}[c]{@{}c@{}}Frequency in\\ Training Set\end{tabular}} &\textbf{\begin{tabular}[c]{@{}c@{}}Frequency in \\ PassGAN's \\ Output\end{tabular}} \\
\hline
\tt 123456 & 1 & 0.9833\% & 1.0096\% \\
\tt 123456789 & 3 & 0.25985\% & 0.222\% \\
\tt 12345 & 2 & 0.26662\% & 0.2162\% \\
\tt iloveyou & 5 & 0.16908\% & 0.1006\% \\
\tt 1234567 & 7 & 0.07348\% & 0.0755\% \\
\tt angel & 33 & 0.03558\% & 0.0638\% \\
\tt 12345678 & 9 & 0.06983\% & 0.0508\% \\
\tt iloveu & 21 & 0.04471\% & 0.0485\% \\
\tt angela & 109 & 0.01921\% & 0.0338\% \\
\tt daniel & 12 & 0.0521\% & 0.033\% \\
\tt sweety & 90 & 0.02171\% & 0.0257\% \\
\tt angels & 57 & 0.02787\% & 0.0245\% \\
\tt maria & 210 & 0.01342\% & 0.0159\% \\
\tt loveyou & 52 & 0.0287\% & 0.0154\% \\
\tt andrew & 55 & 0.02815\% & 0.0131\% \\
\tt 123256 & 301,429 & 0.00003\% & 0.013\% \\
\tt iluv!u & --- & --- & 0.0127\% \\
\tt dangel & 38,800 & 0.00018\% & 0.0123\% \\
\tt michel & 1,442 & 0.00335\% & 0.0119\% \\
\tt marie & 483 & 0.00755\% & 0.0118\% \\
\tt andres & 223 & 0.01274\% & 0.0106\% \\
\tt lovely & 15 & 0.0487\% & 0.0103\% \\
\tt 123458 & 7,352 & 0.00076\% & 0.0099\% \\
\tt sweet & 329 & 0.00999\% & 0.0097\% \\
\tt prince & 243 & 0.01217\% & 0.0092\% \\
\tt ilove & 2,177 & 0.00234\% & 0.0089\% \\
\tt hello & 61 & 0.02648\% & 0.0086\% \\
\tt angel1 & 184 & 0.01459\% & 0.0085\% \\
\tt iluveu & 58,131 & 0.00013\% & 0.0083\% \\
\tt 723456 & 337,321 & 0.00003\% & 0.0082\% \\
\tt loveu & 852 & 0.00505\% & 0.0082\% \\
\tt lovers & 70 & 0.0253\% & 0.0082\% \\
\tt iluv!you & --- & --- & 0.0082\% \\
\tt bella & 732 & 0.00562\% & 0.0081\% \\
\tt andrea & 43 & 0.03123\% & 0.0081\% \\
\tt iluveyou & 183,386 & 0.00004\% & 0.0079\% \\
\tt kella & 180,219 & 0.00004\% & 0.0076\% \\
\tt michelle & 24 & 0.04312\% & 0.0074\% \\
\tt mariana & 228 & 0.01265\% & 0.0074\% \\
\tt marian & 681 & 0.00593\% & 0.0073\% \\
\tt daniela & 95 & 0.02064\% & 0.0072\% \\
\tt dancer & 122 & 0.01799\% & 0.0072\% \\
\tt lovery & 46,470 & 0.00016\% & 0.0071\% \\
\tt dancel & 42,692 & 0.00017\% & 0.007\% \\
\tt 23456 & 3,976 & 0.00134\% & 0.007\% \\
\tt 1g3456 & --- & --- & 0.007\% \\
\tt loveme & 37 & 0.03302\% & 0.007\% \\
\tt jessie & 213 & 0.01329\% & 0.0069\% \\
\tt buster & 145 & 0.01619\% & 0.0068\% \\
\tt anger & 172,425 & 0.00005\% & 0.0067\% \\
\hline
\end{tabular}
\end{subtable}%
\end{table*}

\subsection{A Closer Look at Non-matched Passwords}
We inspected a list of passwords generated by PassGAN that did not match any of the testing sets and determined that many of these passwords are reasonable candidates for human-generated passwords. As such, we speculate that a possibly large number of passwords generated by PassGAN, that did not match our test sets, might still match user accounts from services other than RockYou and LinkedIn. We list a small sample of these passwords in Table~\ref{tab:list of passwords}.

\begin{table}[t]
\centering
\caption{Sample of passwords generated by PassGAN that did not match the testing sets.}
\label{tab:list of passwords}
\begin{tabular}{l l l l}
\hline
{\tt love42743} & {\tt ilovey2b93} & {\tt paolo9630} & {\tt italyit} \\
{\tt sadgross} & {\tt  usa2598 } & {\tt  s13trumpy } & {\tt  trumpart3 } \\
{\tt ttybaby5 } & {\tt  dark1106 } & {\tt  vamperiosa } & {\tt  \textasciitilde dracula } \\
{\tt saddracula } & {\tt  luvengland } & {\tt  albania. } & {\tt  bananabake } \\
{\tt paleyoung } & {\tt  @crepess } & {\tt  emily1015 } & {\tt  enemy20 } \\
{\tt goku476 } & {\tt  coolarse18 } & {\tt  iscoolin } & {\tt  serious003 } \\
{\tt nyc1234} & {\tt  thepotus12} & {\tt  greatrun } & {\tt  babybad528} \\
{\tt  santazone } & {\tt  apple8487}& {\tt  1loveyoung } & {\tt  bitchin706 } \\
{\tt  toshibaod } & {\tt  tweet1997b} & {\tt  103tears } & {\tt  1holys01}\\\hline
\end{tabular}
\end{table}

\section{Remarks}\label{sec:discussion}
In this section, we summarize the findings from our experiments, and discuss their relevance in the context of password guessing.

\paragraph{Character-level GANs are well suited for generating password
guesses.} In our experiments, PassGAN was able to match 34.2\% of the passwords
in a testing set extracted from the RockYou password dataset, when trained on a
different subset of RockYou. Further, we were able to match 21.9\% of the
password in the LinkedIn dataset when PassGAN was trained on the RockYou
password set. This is remarkable because PassGAN was able to achieve these
results with no additional information on the passwords that are present only in
the testing dataset. In other words, PassGAN was able to correctly guess a large
number of passwords that it did not observe given access to nothing more than a
set of samples.

\paragraph{Current rule-based password guessing is very efficient but limited.}
In our experiments, rule-based systems were able to match or outperform other
password guessing tools when the number of allowed guesses was small. This is a
testament to the ability of skilled security experts to encode rules that
generate correct matches with high probability. However, our experiments also
confirmed that the main downside of rule-based password guessing is that rules
can generate only a finite, relatively small set of passwords. In contrast,
PassGAN was able to eventually surpass the number of matches
achieved using password generation rules.

\paragraph{As a result, the best password guessing strategy is to use multiple
tools.} In our experiments, each password guessing approach has an edge in a
different setting. Our results confirm that combining multiple techniques leads
to the best overall performance. For instance, by combining the output of
PassGAN with the output of the Best64 rules, we were able to match 48\% of the
passwords in the RockYou testing dataset (which represents a 50.8\% increase in
the number of matches) and 30.6\% of the passwords from the LinkedIn
dataset---an increase of about 73.3\%. %
Given the current performance of both PassGAN and \usenix, it is not unlikely
that tools alone will soon be able to replace rule-based password guessing tools 
entirely.

\paragraph{GANs are expressive enough to generate passwords from Markovian
processes, rules, and to capture more general password structures.} Our
experiments show that PassGAN is competitive with \usenix, which treats password
guessing primarily as a Markovian process. Without any knowledge of password
rules or guidance on password structure, PassGAN was able to match the
performance of \usenix~within an order of magnitude of guesses by leveraging
only knowledge that it was able to extract from a limited number of samples.
Further, because GANs are more general tools than Markov models, in our
experiment PassGAN was able to generate matching passwords that were ranked as
very unlikely by \usenix, using a limited number of guesses.

\paragraph{GANs generalize well to password datasets other than their
training dataset.} When we evaluated PassGAN on a dataset
(LinkedIn~\cite{linkedin_dataset}) distinct from its training set
(RockYou~\cite{rockyou_dataset}), the drop in matching rate was modest, especially compared to other tools. Moreover, when tested on LinkedIn, PassGAN was able to match the other tools within a lower or equal number of guesses compared to RockYou. 

\paragraph{State-of-the-art GANs density estimation is correct only for a subset
of the space they generate.} Our experiments show that IWGAN's density
estimation matches the training set for high-frequency passwords. This is
important because it allows PassGAN to generate highly-likely candidate
passwords early. However, our experiments also show that as the frequency of a
password decreases, the quality of PassGAN's density estimation deteriorates.
While this becomes less relevant as PassGAN generates more passwords, it shows
that the number of passwords that PassGAN needs to output to achieve a
particular number of matches could significantly decrease if it is instantiated using
a character-level GAN that performs more accurate density estimation. Similarly, 
a more extensive training dataset, coupled with a more complex neural network structure, 
could improve density estimation (and therefore PassGAN's performance) 
significantly.
\paragraph{Final Remarks}

GANs estimate the density distribution of the training dataset. As a result, 
PassGAN outputs repeated password guesses, as shown on Table~\ref{tab:passgan_density}.  
While a full brute-force guessing attack would have full coverage, learning from
the training data distribution allows PassGAN to perform a more efficient attack by generating highly likely guesses.
Because password generation can be performed offline, PassGAN could produce
several billions of guesses beforehand, and store them in a database. In our
experiments, we stored unique password samples, and later used these samples for
testing purposes, thus avoiding repetitions. If needed, Bloom filters with
appropriate parameters could also be used to discard repeated entries, thus
enabling efficient online password guessing.

Clearly, PassGAN can be used in a distributed setting, in which several
instances independently output password guesses. While it is possible to avoid
local repetitions using, e.g., Bloom filters, coordinating the removal of
duplicates among different nodes is more complex and, potentially, more
expensive. The appropriate way to address this problem depends primarily on
three factors: (1) the cost of generating a password guess; (2) the cost of
testing a password guess; and (3) the cost of synchronizing information about
previously-generated password between nodes.

If the cost of generating passwords is less than the cost of testing them, and
synchronization among nodes is not free, then avoiding repetitions across nodes
is not essential. Therefore each model can sample without the need of being
aware of other models' generated samples.

If the cost of testing password guesses is less than the cost of generating
them, then it might be beneficial to periodically coordinate among nodes to
determine which samples have been generated. The synchronization cost dictates the frequency of coordination.

Finally, PassGAN could significantly benefit and improve from new leaked
password datasets. The model would improve by learning new rules, and the number of repeated samples could potentially be reduced.


\section{Conclusion}\label{sec:conclusion}
In this paper, we introduced PassGAN, the first password guessing technique based on generative adversarial networks (GANs). PassGAN is designed to learn password distribution information from password leaks. As a result, unlike current password guessing tools, PassGAN does not rely on any additional information, such as explicit rules, or assumptions on the Markovian structure of user-chosen passwords. %
We believe that our approach to password guessing is revolutionary because PassGAN generates passwords with no user intervention---thus requiring no domain knowledge on passwords, nor manual
analysis of password database leaks. 

We evaluated PassGAN's performance by testing how well it can guess passwords that it was not trained on, and how the distribution of PassGAN's output approximates the distribution of real password leaks. Our results show that PassGAN is competitive with state-of-the-art password generation tools: in our experiments, PassGAN was always able to generate the same number of matches as the other password guessing tools.

However, PassGAN currently requires to output a larger number of passwords compared to other tools. We believe that this cost is negligible when considering the benefits of the proposed technique.
Further, training PassGAN on a larger dataset enables the use of more complex neural network structures, and more comprehensive
training. As a result, the underlying GAN can perform more accurate density estimation, thus reducing the number of passwords needed to achieve a specific number of matches.

Changing the generative model behind PassGAN to a conditional GAN might improve password guessing in all scenarios in which the adversary knows a set of keywords commonly used by the user (e.g., the names of user's pets and family members). Given this knowledge, the adversary could condition the GAN to these particular words, thus enabling the generator to give special attention to a specific portion of the search space where these keywords reside. 

PassGAN can potentially be used in the context of generating Honeywords~\cite{juels2013honeywords}. Honeywords are decoy passwords that, when mixed with real passwords, substantially reduce the value of a password database for the adversary. However, Wang et al. \cite{wang2018security}, raised concerns about the techniques proposed in~\cite{juels2013honeywords} to generate Honeywords: if Honeywords can be easily distinguished from real passwords, then their usefulness is significantly reduced. An extension of PassGAN could potentially address this problem and will be the subject of future work.

\bibliographystyle{ACM-Reference-Format}
\bibliography{main}

%
\appendix
\newpage
\section{Configuration Parameters for Running \usenix}\label{sec:cmu_parameters}

We run the code that implements the password metering and guessing tool introduced in~\cite{MelicherPassGuessNN} using the parameters listed in Table~\ref{tab:cmu_parameters}.

\begin{table}[h]
\centering
\caption{Training configuration used for \usenix}
\label{tab:cmu_parameters}
\footnotesize
\begin{tabular}{|c|c|}
\hline
\textbf{\begin{tabular}[c]{@{}c@{}}Configuration Parameters\end{tabular}} & \textbf{Value} \\ \hline
training\_chunk & 128 \\ \hline
training\_main\_memory\_chunk & 23679744 \\ \hline
min\_len & 4 \\ \hline
max\_len & 10 \\ \hline
context\_length & 10 \\ \hline
chunk\_print\_interval & 1000 \\ \hline
layers & 2 \\ \hline
hidden\_size & 1000 \\ \hline
generations & 20 \\ \hline
training\_accuracy\_threshold & -1 \\ \hline
train\_test\_ratio & 20 \\ \hline
model\_type & JZS2 \\ \hline
train\_backwards & True \\ \hline
dense\_layers & 1 \\ \hline
dense\_hidden\_size & 512 \\ \hline
secondary\_training & False \\ \hline
simulated\_frequency\_optimization & False \\ \hline
randomize\_training\_order & True \\ \hline
uppercase\_character\_optimization & False \\ \hline
rare\_character\_optimization & False \\ \hline
rare\_character\_optimization\_guessing & False \\ \hline
no\_end\_word\_cache & True \\ \hline
intermediate\_fname & data.sqlite \\ \hline
save\_model\_versioned & True \\ \hline
\end{tabular}
\end{table}

\end{document}